\documentclass[12pt, margin=1in, reqno]{article}
	\usepackage[left=1in, top=1in, right=1in, paperwidth=8.5in, paperheight=11in, textheight=9in]{geometry} 
	\usepackage{graphicx, color, soul, wrapfig, array, subcaption, mdframed, float, enumitem, amsmath, amssymb, courier} 
	\usepackage[font=footnotesize]{caption}
	\usepackage[outdir=./]{epstopdf}
	\usepackage{setspace}		
	\usepackage[superscript]{cite}
	\usepackage[tiny,compact]{titlesec}
	\titlespacing{\section}{0pt}{*0}{*0}
	\titlespacing{\subsection}{0pt}{*0}{*0}
	\titlespacing{\subsubsection}{0pt}{*0}{*0}

\begin{document}

{  \begin{center}  
	\large{ \bf{Distinguishing failure modes at the molecular level by examining heterogeneity in local structures and dynamics}}  \\
	\vspace{3mm}
	\normalsize{Emily Y. Lin, Robert A. Riggleman} \\
	\normalsize{University of Pennsylvania, Philadelphia, PA} 
\end{center} }

\doublespacing
\section{Abstract}

Brittle failure is ubiquitous in amorphous materials that are sufficiently cooled below their glass transition temperature, $T_g$. This catastrophic failure mode is limiting for amorphous materials in many applications, and many fundamental questions surrounding it remain poorly understood. Two challenges that prevent a more fundamental understanding of the transition between a ductile response at temperatures near $T_g$ to brittle failure at lower temperatures are i) a lack of computationally inexpensive molecular models that capture the failure modes observed in experiments and ii) the lack of quantitative metrics that can distinguish various failure mechanisms. In this work, we use molecular dynamics simulations to capture ductile-to-brittle transition in glass-forming short-chain polymer systems by using a modified Lennard-Jones potential to describe non-bonded interactions between the monomers. We characterized the effects of this new potential on macroscopic mechanical properties as well as microscopic structural and dynamical differences during deformation. Lastly, we present  quantitative metrics that distinguish between different failure modes.


\section{Introduction}

Glass-forming polymers play a crucial role in many applications, and amorphous polymers remain one of the few glass-forming materials commonly found in applications. Other families of glasses, such as metallic glasses, are too brittle for many potential applications, and understanding the dynamics in these systems during deformation has become an active topic of research in the last few decades. Glassy materials tend to fail catastrophically in a brittle fashion under high loading conditions, which can be problematic in applications where the structural integrity of these materials is critical to their function\cite{Hufnagel-alloy, Chen-metallic-glass, Ritchie2011}. In crystalline systems, defects propagate during deformation to provide ductility, and the motion of the defects can be easily monitored. However, in disordered solids defects cannot be easily identified, and particle rearrangements that lead to failure become much more complicated to parse. Plastic events are present even when the macroscopic properties indicate the material is still in the elastic regime of the deformation, and they eventually organize into the shear band where the material ultimately fails.\cite{Falk1998d2min, Langer2004, FalkLanger2011, Barrat2006-plastic-response, RAR-PRE} 

Failure in non-polymeric disordered solids after yielding can take several forms. For ductile materials, the shear band draws down to a neck near the yield, at which the sample breaks after significant plasticity. For brittle materials, this necking behavior is not observed; instead, cavities form in and near the shear band, and the sample catastrophically fails via fracture with little additional strain. The transition between ductile and brittle behavior can be tuned through both sample preparation protocols and experimental conditions, such as temperature at which deformation occurs and the strain rate, and because glasses are non-equilibrium systems, the vitrification process and the age of the specimen can also play a role in the failure mechanism.\cite{Johnson2003-MG-strain-rate, Shavit2014, Toepperwein-dePablo2011, Falk2013-cavitation, ShiFalk2006-quench-rate, Ma2007-MG-ductility, 2009strain-rate, DeformationBook, LeeEdiger2010-aging, Arnold1995-aging, Hill1990-aging, RAR2009-glass-creep-mobility} 

At low temperatures, brittle failure is expected for essentially all amorphous solids, including glassy polymers. However, previous studies using the common coarse-grained model polymer with Lennard-Jones (LJ) interaction sites connected by rigid bonds have not observed brittle failure, even for small chain lengths.\cite{Shavit2014} While many of the phenomena produced via this model agree qualitatively very well with experiments\cite{RAR2009-glass-creep-mobility, RAR-2010-SoftMatter}, the inability to capture a transition from homogeneous plastic flow to brittle failure modes remains a significant shortcoming. Previous studies of cavitation and crazing have required imposing triaxial loading\cite{Robbins2001-crazing, RottlerRobbins2002, RottlerRobbins2003-crazes, Toepperwein-dePablo2011, Falk2013-cavitation, RobbinsGrest2016-crazing}; however, this method increases the volume of the specimen, and cannot be used to observe the transition between cavitation and crazing to homogeneous plastic flow. Even the common binary monomer Lennard-Jones glass, which is nominally a model for a metallic glass\cite{Kob-Andersen} apparently does not exhibit the catastrophic failure observed expected at low temperatures\cite{Shi-2010-shear-band-size}. Clearly an inexpensive molecular model, or family of models, that exhibit the expected failure modes would provide a platform to study failure in amorphous solids.

Another challenge facing molecular simulation studies of failure in amorphous solids is the lack of quantitative metrics that can distinguish the various failure modes. Since simulations typically use relatively small sample sizes (with notable exceptions\cite{Metallic-glass-nanowire-BDT, Srolovitz-Greer-2014}) where the failure is not as sudden as in macroscopic samples, the strain to failure is often of limited utility. This situation often leads to describing transitions between various failure modes in a qualitative manner and often through visual inspection.

In this work, we take a first step towards addressing these two issues by characterizing a series of model glass-forming polymers. Beginning with the standard Lennard-Jones models, we systematically modify the temperature, cooling rate used to prepare the glass, and the form of the potential (following ideas proposed in previous work\cite{Falk1999mLJ, Trombach2018}) and study the failure mode of nanopillars under tensile loading. Across temperature and the potentials considered, we observe a spectrum of failure modes including brittle fracture, ductile necking, homogeneous plastic flow, and several intermediate modes. In addition, we develop and examine some quantitative metrics that distinguish the various failure modes. In general, we find that the transitions between the different failure modes are gradual as the system parameters are changed, and we demonstrate that surface roughness alone does not capture the changes observed.

\section{Simulation details}

\begin{figure}[H]
			\centering
				\includegraphics[scale = 0.3]{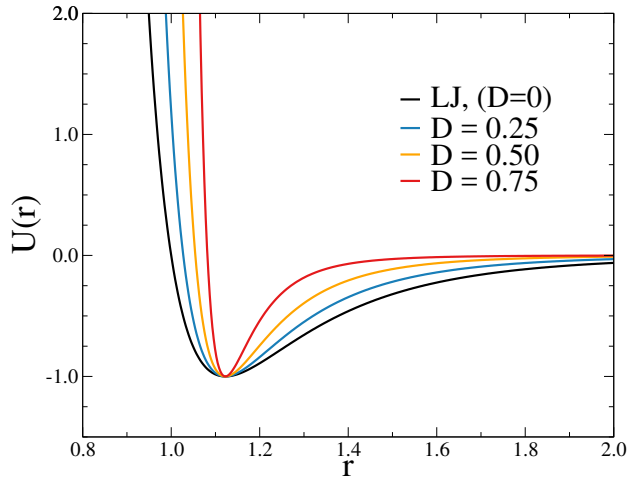}
			\caption{\footnotesize{Standard and modified LJ pair potentials.}}
			\label{fig:pot-fig}
		\end{figure}

For our Molecular Dynamics (MD) simulation studies, we used coarse grained model oligomers with chain length of $N=5$ monomers per chain. The non-bonded interactions take the form of 
\begin{eqnarray}
	U_{ij}^{nb} = 4\epsilon \left[ \left( \frac{\sigma'}{r_{ij} - D} \right)^{12} -  \left( \frac{\sigma'}{r_{ij} - D} \right)^{6} \right] - u_{cut}
\end{eqnarray}
where $D$ is a parameter that allows us to modify the range and curvature at the minimum in our potential, and when $D = 0$ this reduces to the standard LJ potential and has been well established previously\cite{KremerGrest, Shavit2013-thin-film, RAR-Douglas-dePablo2007}. We adjust $\sigma ^{'}=1-D/2^{1/6}$ to fix the location of the minimum, as shown in Figure \ref{fig:pot-fig} for the four values of $D$ used in this study. The bonded interaction in our simulations is a stiff harmonic bonding potential, $U_{ij}^b = (k_h/2)(r_{ij} - \sigma)^2$, where $k_h = 2000 \; \epsilon/\sigma^2$. All units reported in this study are in LJ reduced units: temperature $T = kT^*/\epsilon$, and time $\tau_{LJ} = t^*\sqrt{\epsilon/m\sigma^2}$, where $m$ represents the mass of a single LJ interaction site, $T^*$ and $t^*$ represent temperature and time measured in laboratory units, and $\epsilon$ and $\sigma$ are the parameters of the standard LJ potential ($D=0$). We used the LAMMPS  simulation package with a small simulation time step of $\delta t = 0.000663652 ~~\tau_{LJ} / \rm timestep$\cite{lammps}. We chose this $\delta t$ value to be commensurate with the increased curvature of the modified LJ potential (mLJ), and the mLJ potential is implemented in LAMMPS as the ``lj/expand'' pair style due to its common use in simulating nanoparticles and colloids.

\subsection{Characterization of the mLJ potentials}
		
\begin{figure}[H]
			\centering
				\includegraphics[width = 0.45\columnwidth]{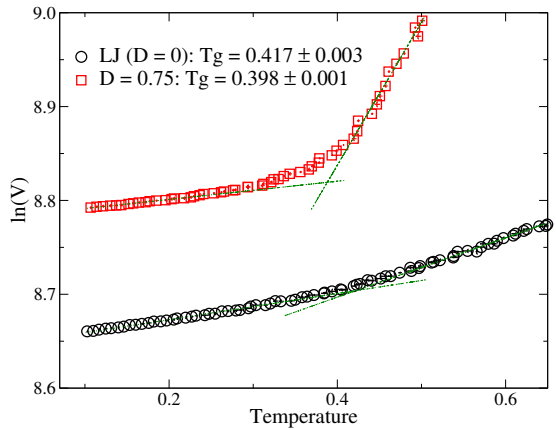}
			\caption{\footnotesize{Characterization of LJ and mLJ ($D = 0.75$) systems. (a). Cooling profiles of LJ and mLJ bulk systems using a quench rate of $\dot{\Gamma} = 1\times10^{-3}$. The dotted lines are linear regression fits used to identify $T_g$.}}
			\label{fig:modTg}
		\end{figure}

For comparing the $T_g$ of the standard and of the modified LJ systems, bulk polymer glass simulations were used. Monodispersed systems consisting of 6000 monomers total were used for LJ and mLJ ($D=0.75$) potentials. Three independent configurations were generated by equilibrating at high temperature for different lengths of time, and uncertainties are taken as standard error calculated using these three configurations. After equilibration at $T \gg T_g$, we rapidly quenched the systems to $T = 0.05 $ at a cooling rate of $\dot{\Gamma} = \Delta T / \Delta t = 1\times10^{-3}$. We collected volume change during cooling, which allowed us to obtain $T_g$ by locating the temperature at which the thermal expansion coefficient $\alpha_T = (\partial \ln v / \partial T)_P$ changes, shown in Figure \ref{fig:modTg}. The $T_g$ for the LJ systems is $0.417\pm0.003$, while the $T_g$ for the mLJ ($D = 0.75$) systems shows a slight decrease at $0.398\pm0.001$. The ratios of thermal expansivity ($\alpha_{T, l} / \alpha_{T, g}$) for the LJ and the mLJ systems are $2.245 \pm 0.001$ and $15.70 \pm 0.09$, respectively. While this quantity for the LJ system matches well with experimental data of a typical polymer, the value for the mLJ system is significantly higher due to the large $\alpha_T$ for the supercooled liquid. However, since our interest in this study is mainly to investigate the mechanics of the glass pillar failure mechanism and the transition from necking to fracture failure at constant temperature below $T_g$, properties of the supercooled liquid do not play a significant role in our simulations. 

\subsection{Strain localization studies}

Cylindrical pillars with a diameter of approximately $30\sigma$ and an aspect ratio of approximately 2 were generated; this diameter is large enough to allow for bulk like dynamics in the center of the pillar \cite{Shavit2014}. Four independent configurations of pillars for each $D$ value were generated at high temperatures and equilibrated to erase previous thermal history. Pillars were equilibrated in the NVT ensemble at $T \gg T_g$ with periodic boundary condition along the length of the pillars, then quenched to $T= 0.05$. A repulsive wall potential was used at high temperature to maintain pillar geometry. Configurations at a total of six temperatures below $T_g$ ($T \in [0.05, 0.3]$ with an increment of 0.05) were collected during cooling in order to ensure a consistent sample history. We deformed each pillar at constant temperature and constant true strain rate ($\dot{\epsilon} = 1\times10^{-4}$) by applying uniaxial tension along the z direction.

Using particle configurations during deformation, we calculated the local strain rate associated with each particle\cite{JenR2015J2, Falk1998d2min} 
\begin{eqnarray}
J_2(\epsilon, \epsilon + \Delta \epsilon) 
= 
\frac{1}{\Delta \epsilon} 
\sqrt{ \frac{1}{d} Tr \left[ \frac{1}{d} (\mathbf{J}_i^T \mathbf{J} - \mathbf{I}) - \frac{1}{d} Tr(\mathbf{J}_i^T \mathbf{J} - \mathbf{I}) \right]^2 }
\end{eqnarray}
where $d$ is the dimensionality of the system, $\mathbf{J}_i$ is the best affine transformation matrix for particle i at strain $\epsilon$, given a lag strain of $\Delta \epsilon$, and $\textbf{I}$ is the identity matrix. This local strain rate is calculated for each monomer by calculating the best-fit local affine transformation matrix, constructing the Lagrangian strain tensor, and extracting the deviatoric components. Particles with large $J_2$ values have a higher non-affine strain rate in their local environment. 

While it is informative to see the variation of $J_2$ for each monomer, we also develop metrics to characterize the mesoscale nature of the strain response in order to distinguish strain localization and homogeneous plastic flow. We divided the cylindrical pillar axially into 20 slabs of equal thickness, and calculated the spatial fluctuations of average $J_2$ values between pairs of slabs. We define $S_L$ as this quantity averaged over all pairs of slabs,
		\begin{equation}
			S_L (T, \epsilon; \epsilon_w) = \frac{1}{n_b(n_b - 1)} \sum_{i=1}^{n_b} \sum_{j=1, j \neq i}^{n_b} \langle (\overline{J_2(i,\epsilon)} - \langle \overline{J_2(j)} \rangle_{\epsilon_w} )^2 \rangle_{\epsilon_w} \cdot \langle (\overline{J_2(j,\epsilon)} - \langle \overline{J_2(i)} \rangle_{\epsilon_w} )^2 \rangle_{\epsilon_w}
			\label{Sl}
		\end{equation}
where $n_b$ denotes the number of blocks, the overhead bar represents averaging over all particle $J_2$ values within the block, $\epsilon_w$ represents the strain window over which we averaged $J_2$ values in addition to slab average. In all of our calculations, we chose $\epsilon_w$ to be the same as the lag strain used in calculations of $J_2$, $\Delta \epsilon = 1\%$. This additional average over strain window allows us to focus on spatial fluctuations that are relatively long lived. When the strain is homogeneous, the averages of $\overline{J_2}$ in any two slabs $i$ and $j$ will be approximately the same, and as a result $\langle (\overline{J_2(i,\epsilon)} - \langle \overline{J_2(j)} \rangle_{\epsilon_w}$ will be nearly zero and $S_L$ is small. In contrast, when there are large spatial variations in the strain rate, $S_L$ increases sharply, indicating strain localization. Each slab is approximately 5 monomers thick prior to deformation, and increases in thickness affinely as strain increases.
		
\subsection{Voronoi volume calculations}

We used a C++ library (Voro++) to perform Voronoi tessellation of our nanopillar, and to compute the resulting Voronoi volumes.\cite{voro++} For the tessellation routine, we treated the center of our monomer as the centroid of the Voronoi cell. After the calculation of the Voronoi volumes for all particles in our nanopillars for all strains, we discard the particles that are on the surfaces of the pillars by discounting any particle that has a Voronoi volume larger than $8 \sigma^3$. We examined the distributions of Voronoi volumes before and after strain is applied for different $D$ values and temperatures. Finally, to identify particles near a fracture surface, we sorted all of the non-surface particle Voronoi volumes by their magnitude, and defined a cutoff Voronoi volume for each $D$ value and $\epsilon$ to be the 90\textsuperscript{th} percentile of the Voronoi volumes. 
		
\section{Results}

\subsection{Macroscopic Mechanical Response}

\begin{figure}[H]
			\centering
			\begin{subfigure}{0.45\columnwidth}
				\includegraphics[width = \linewidth]{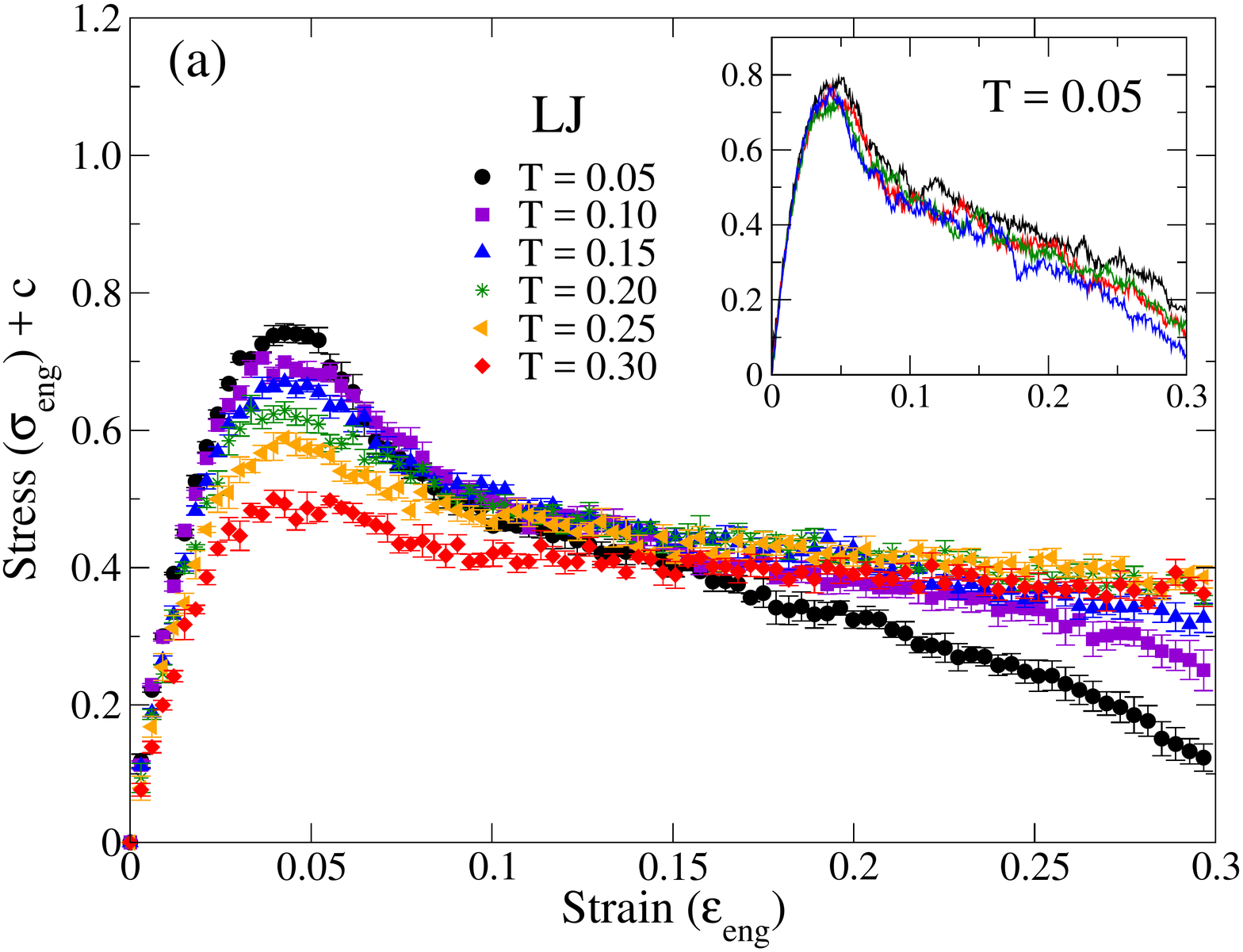}
			\end{subfigure} 
			~
			\begin{subfigure}{0.45\columnwidth}
				\includegraphics[width = \linewidth]{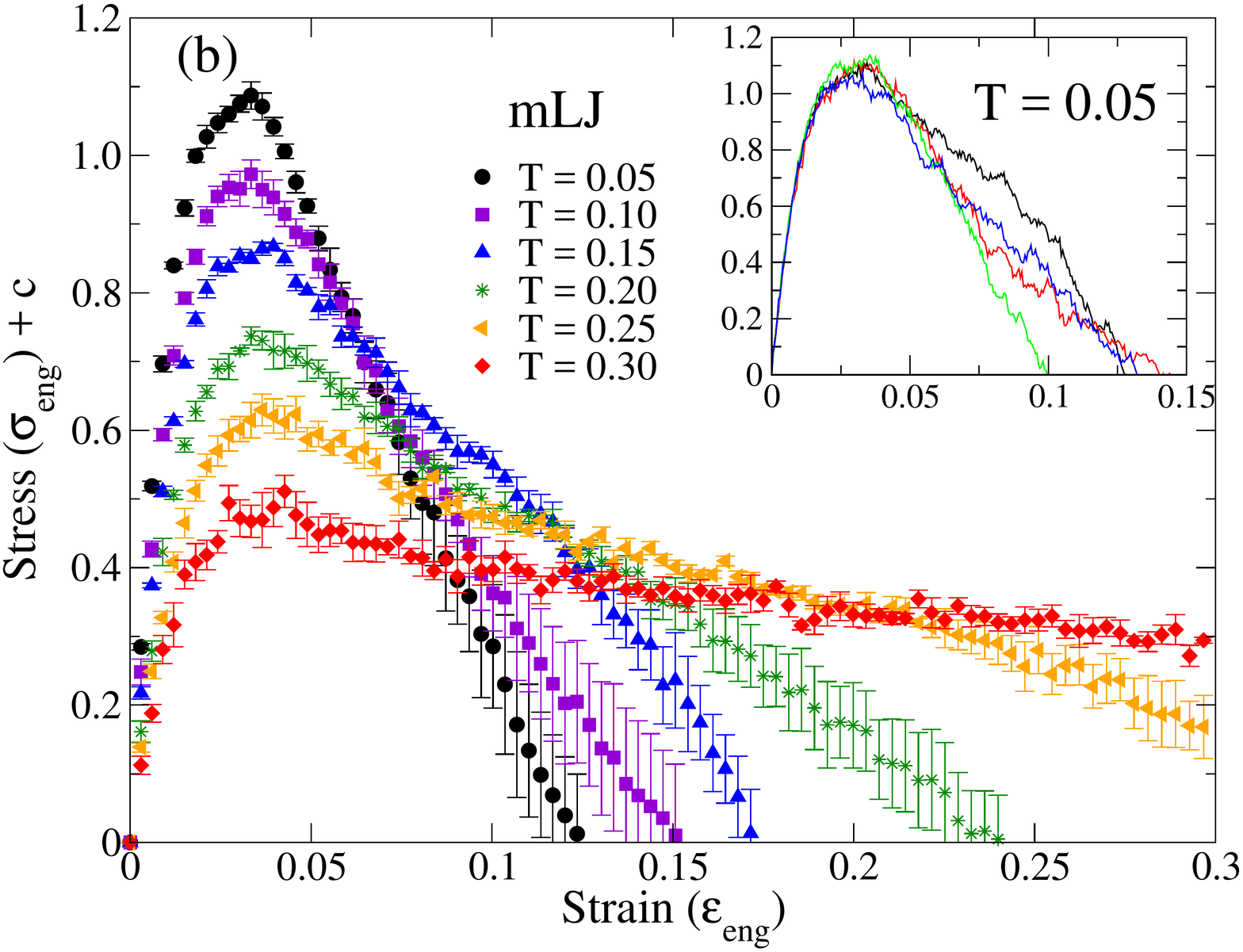}
			\end{subfigure}
			\\
			\begin{subfigure}{0.45\columnwidth}
				\includegraphics[width = \linewidth]{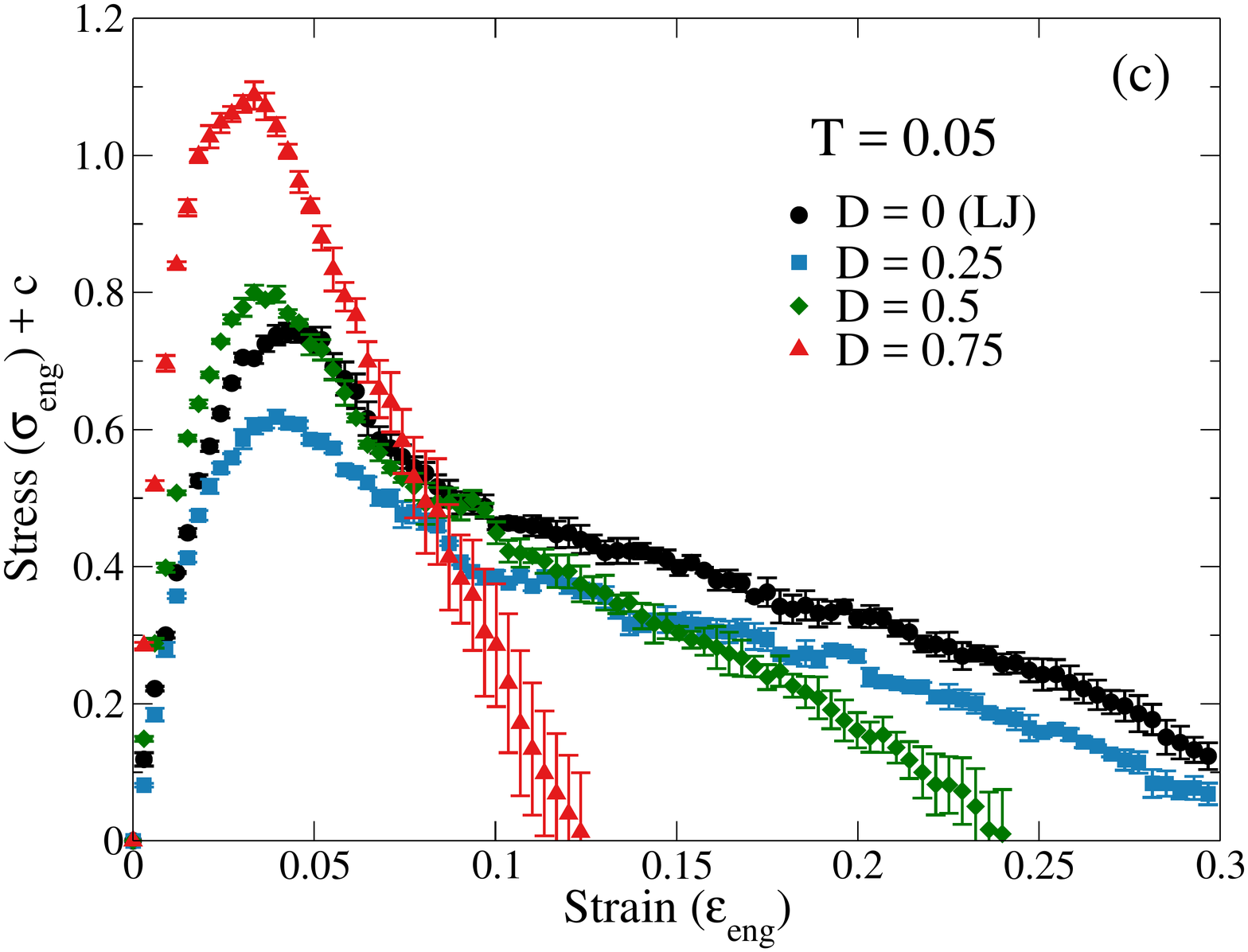}
			\end{subfigure}
			~
			\begin{subfigure}{0.45\columnwidth}
				\includegraphics[width = \linewidth]{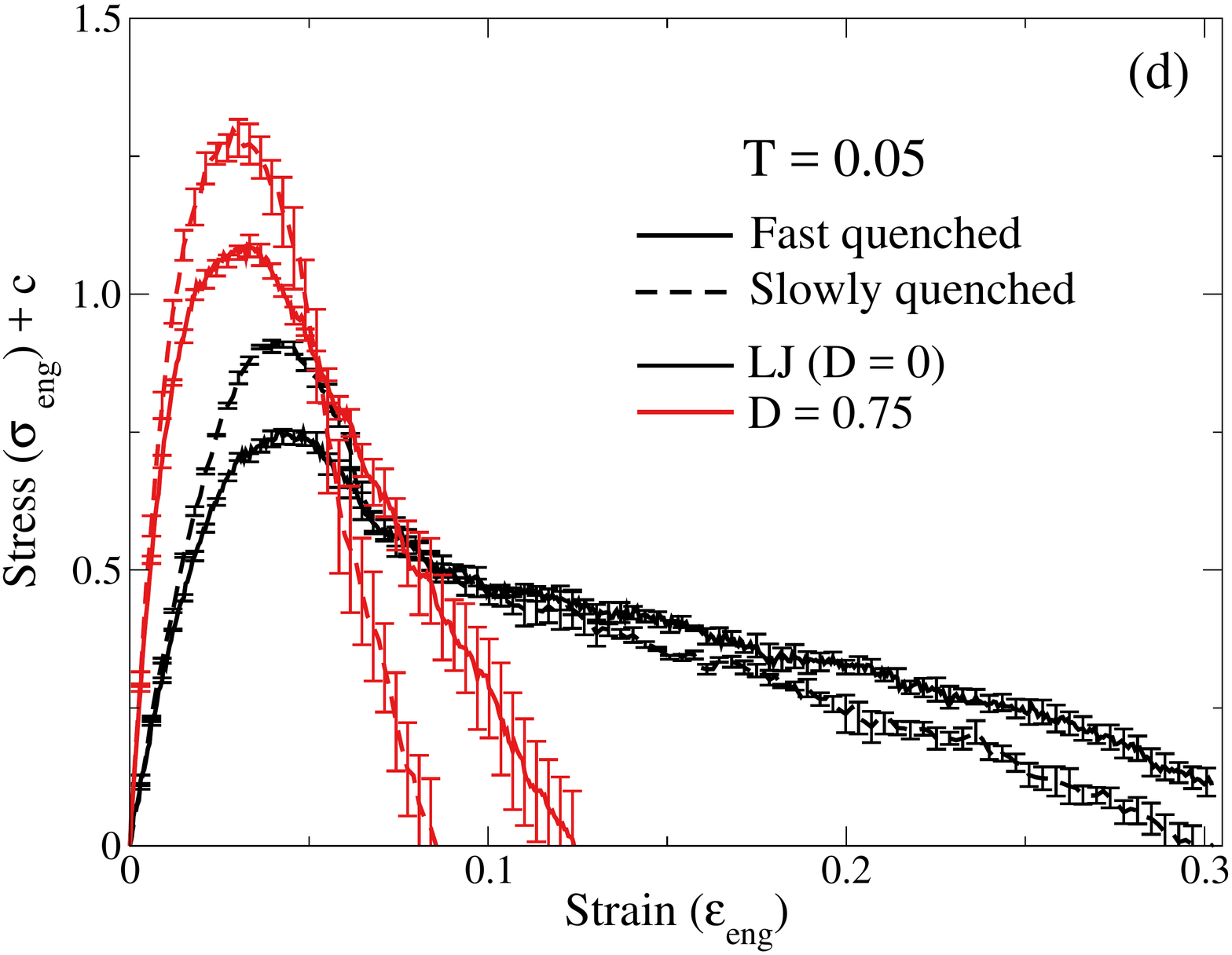}			
			\end{subfigure}			
			\caption{\footnotesize{Stress-strain curve of (a) LJ and (b) mLJ pillars at six temperatures below $T_g$. Pillars at temperatures below $T_g$ were deformed uniaxially at a true strain rate of $\dot{\epsilon} = 1\times 10^{-4}$. Insets show variations amongst four independent pillar configurations at the lowest temperature. (c). Pillars at T =0.05 using all four values of D (0.00, 0.25, 0.50, 0.75). (d). Comparison between two quench rates for the LJ and mLJ (D = 0.75) pillars at T = 0.05. A stress offset is used by subtracting the initial system stress (constant c) to start all curves at $\sigma_{eng} = 0$}}
			\label{fig:stress}
		\end{figure}

We began our analyses by comparing the mechanical response of the LJ and mLJ (with $D = 0.75$) systems. First, we used engineering stress data collected during our deformation simulations to plot the stress-strain relationship for both systems to investigate the effect of modifying the LJ potential on the yield strength and the elastic modulus. As shown in Fig. \ref{fig:stress}, both the yield strength and the elastic modulus are higher in the mLJ system, and they decrease modestly with increasing temperature, as expected. The effects of temperature are more pronounced in the mLJ system with $D=0.75$. At the highest temperature in our study, the LJ and mLJ pillars exhibit similar stress-strain behavior, suggesting that the effect of modifying interaction potential on bulk mechanical properties becomes minimal at higher temperatures close to $T_g$. The differences in behavior at low temperature are stark: LJ exhibits significant plasticity, while mLJ stress curve decreases sharply after yielding at $\epsilon \approx 4\%$. In Figure \ref{fig:stress}c, we show the stress-strain relationship obtained at the lowest temperature for systems with each $D$ considered; as $D$ increases, the stress-strain curve shows less plasticity after yield. 

The effect of quench rate exhibits the expected behaviors as quench rate decreases. To examine the effect of quench rate, we constructed pillars using quench rate two orders of magnitude slower ($\dot{\Gamma} = 1\times10^{-5}$) than the fast quenched samples. Figure \ref{fig:stress}d demonstrates that the slowly quenched pillars at a given $D$ and temperature tend to have higher yield stress and higher elastic modulus than their fast quenched counterparts, consistent with the common observation that physical aging of a glass increases its modulus.

\begin{figure}[H]
			\centering
			\includegraphics[scale = 0.3]{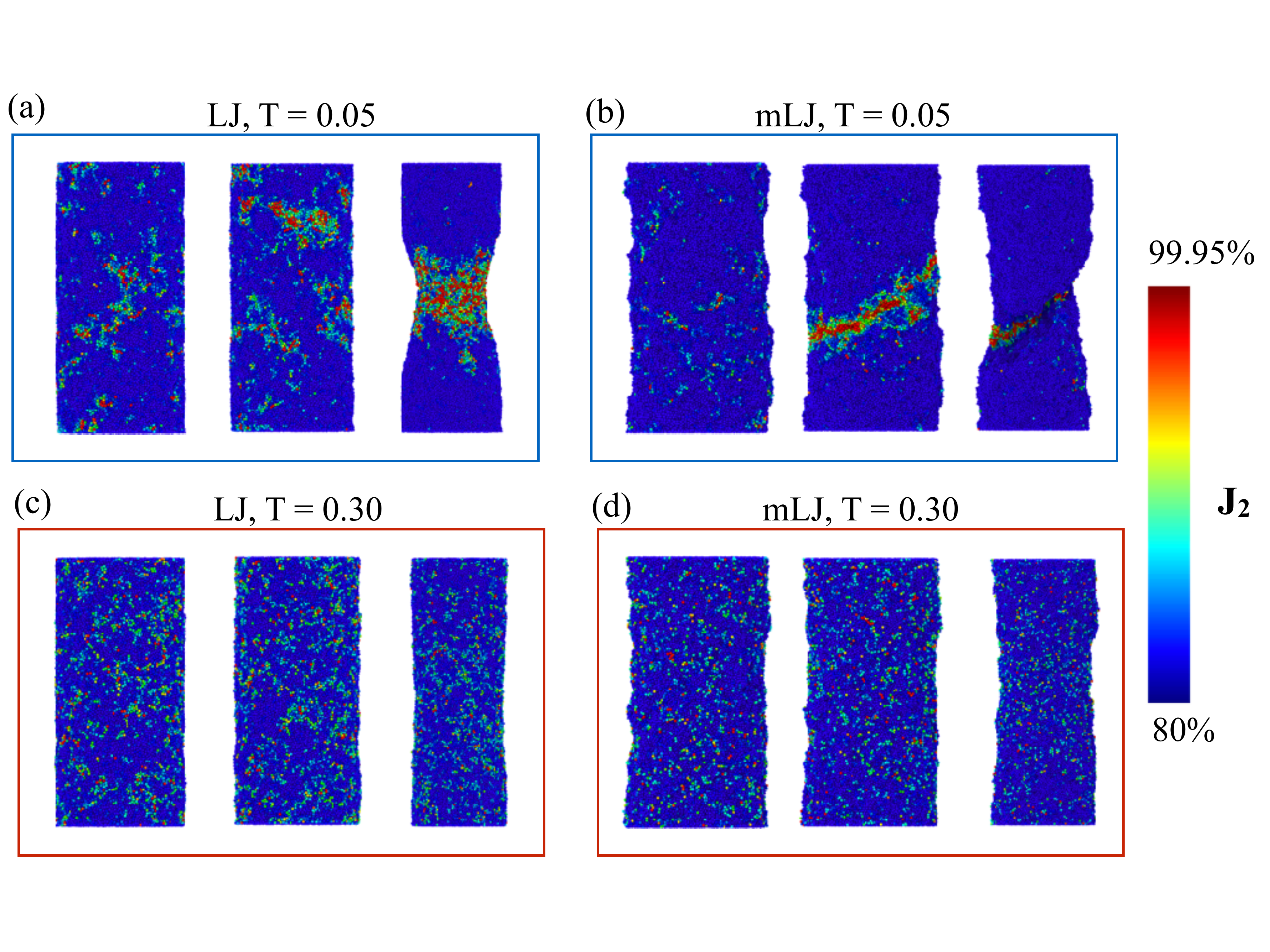}
			\caption{\footnotesize{Visualization of the strain field of a subset of our polymer pillars. Pillars are colored with a selected range of $J_2$ values in each frame (80th to 99.5th percentiles). The four framed sets show (a) LJ at T = 0.05, (b) mLJ at T = 0.05, (c) LJ at T = 0.30, and (d) mLJ at T = 0.30. Within each set, we visualized the strain field on the pillars in the elastic (left: $\epsilon \approx 2.5\%$}), yield (middle: $\epsilon \approx 5.0\%$), and post-yield (right: $\epsilon \approx 28\%$) regimes.}
			\label{fig:pillars}
		\end{figure}

Figure \ref{fig:pillars} demonstrates the difference in failure mechanism between the two potentials with $D=0$ and $0.75$. Snapshots of our simulation were taken while the deformation is in elastic, yield, and post-yield regimes, and instantaneous $J_2$ values calculated over a lag strain of 1\% are employed to color each monomer. The color scale is compressed so that blue represents 80th percentile and lower, and red represents 99.95th percentile and higher, and these percentiles are defined based on $J_2$ data collected from the entire simulation. Each framed (blue and red) set represents a configuration at the stated conditions: LJ vs. mLJ ($D = 0.75$), and $T = 0.05$ vs. $T = 0.30$. The LJ system exhibits necking at the lowest temperature in our study, implying the system retains significant ductility even far below its $T_g$, while the mLJ system at low temperature exhibits brittle failure with a fracture surface at approximately $45^{\rm o}$ to the normal. At $T = 0.30$, both potentials exhibit homogeneous plastic flow. 	
	
\subsection{Microscopic response}

\begin{figure}[H]
			\centering
				\includegraphics[scale = 0.5]{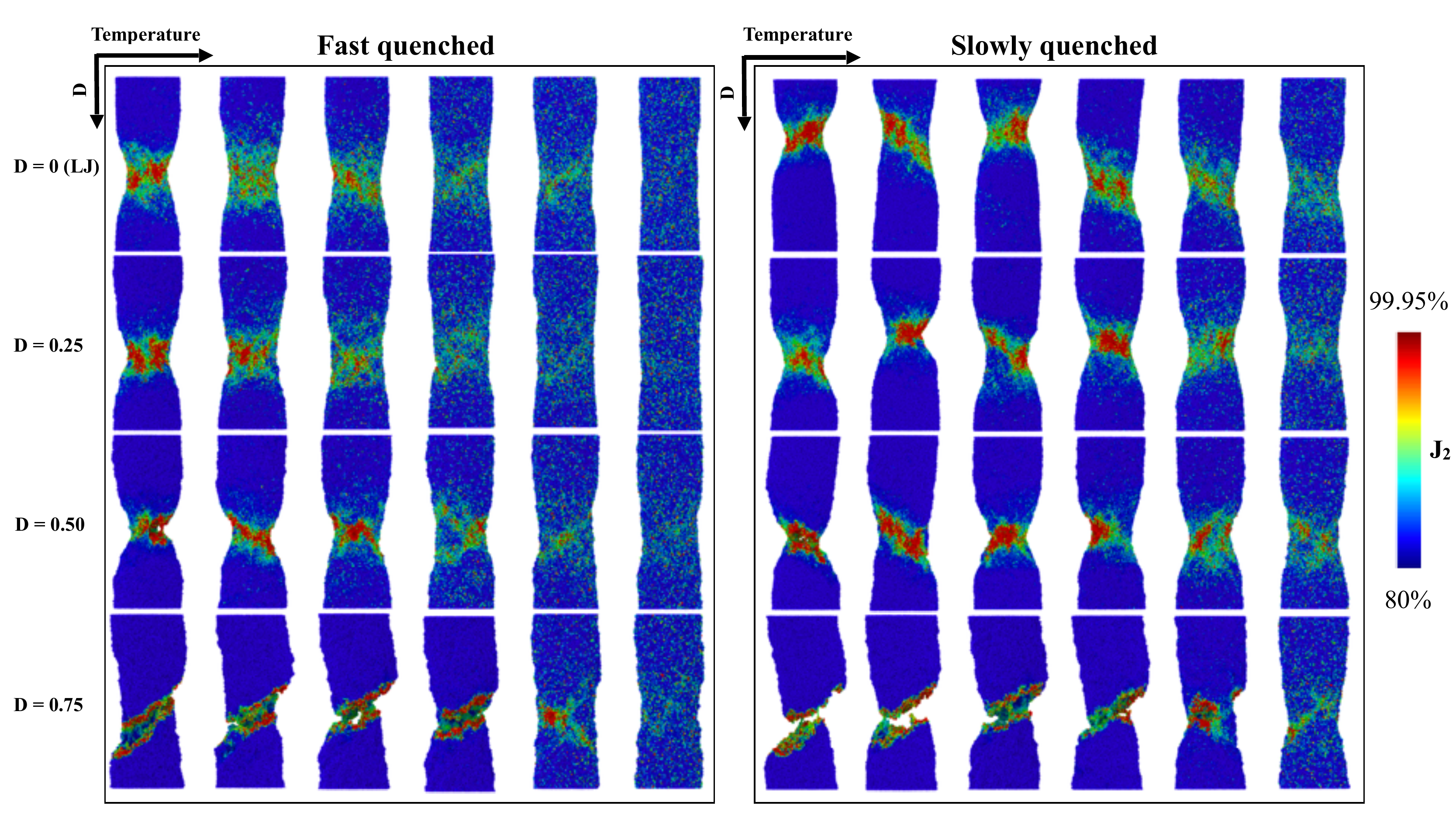}
			\caption{\footnotesize{Representative images of pillars at the end of our deformation simulations ($\epsilon \approx 30\%$. The particles are highlighted using cumulative $J_2$ values calculated comparing the configurations at the initial and the end of the entire simulation. Particle color scales are determined using the 80th and 99.95th percentile of all $J_2$ values for the cutoff for blue and red colors respectively. This means the blue particles have lowest cumulative $J_2$ values and the red particles have highest cumulative $J_2$ values. In each of the cooling rates, temperature increases from left to right, and D value increases from top to bottom. }}
			\label{fig:final-pics}
		\end{figure}

Figure \ref{fig:final-pics} visualizes the post-yield strain field as a function of temperature for all of the potentials considered. We can see that for a given D value and quench rate, the degree of strain localization decreases and eventually vanishes as temperature increases. For the same quench rate and temperature combination, as we increase D we transition from necking deformation to brittle fracture failure modes. Additionally, the slowly quenched pillars have more pronounced strain localization than their fast quenched counterparts, especially at higher temperatures.   

\begin{figure}[H]
			\centering
			\begin{subfigure}{0.45\columnwidth}
				\includegraphics[width = \linewidth]{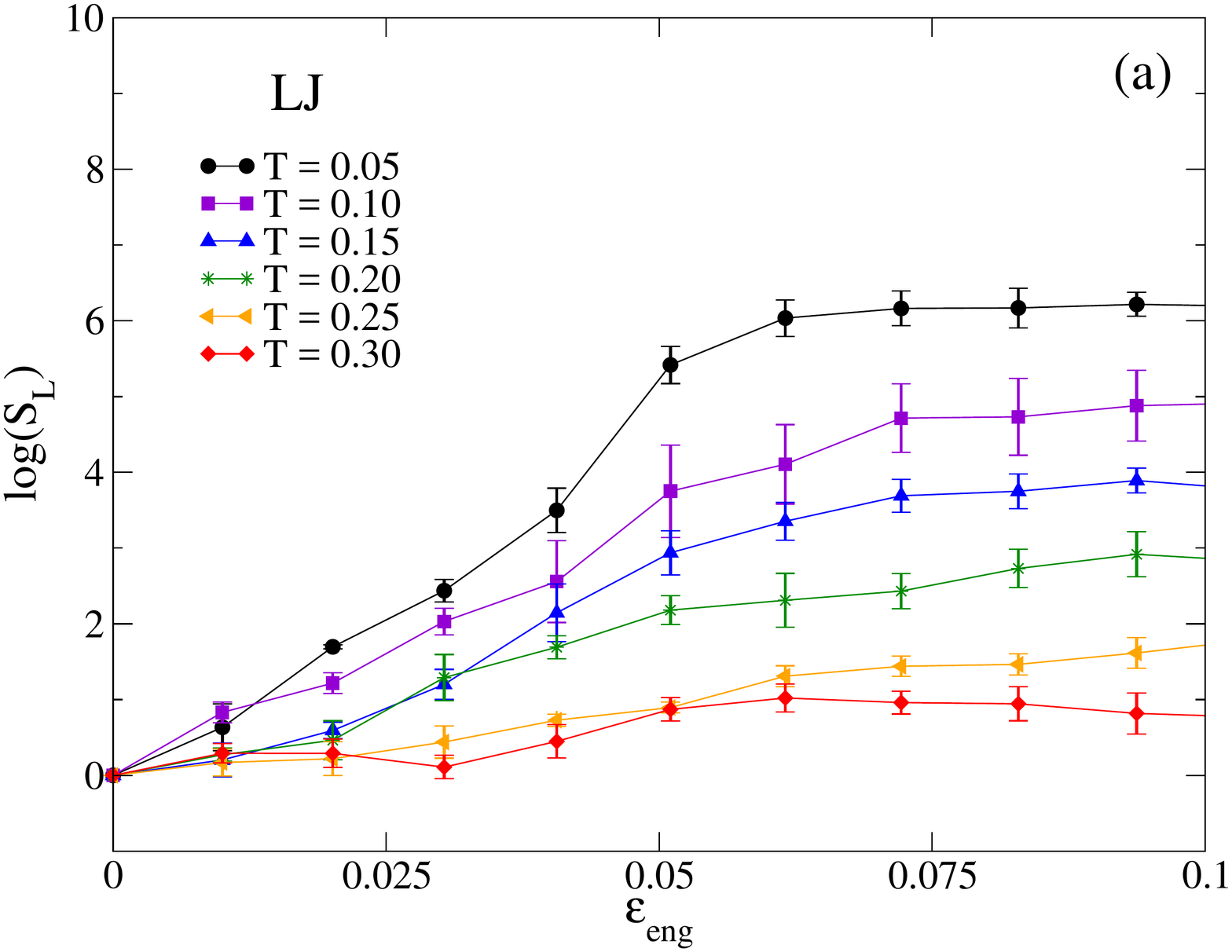}
			\end{subfigure} 
			~
			\begin{subfigure}{0.45\columnwidth}
				\includegraphics[width = \linewidth]{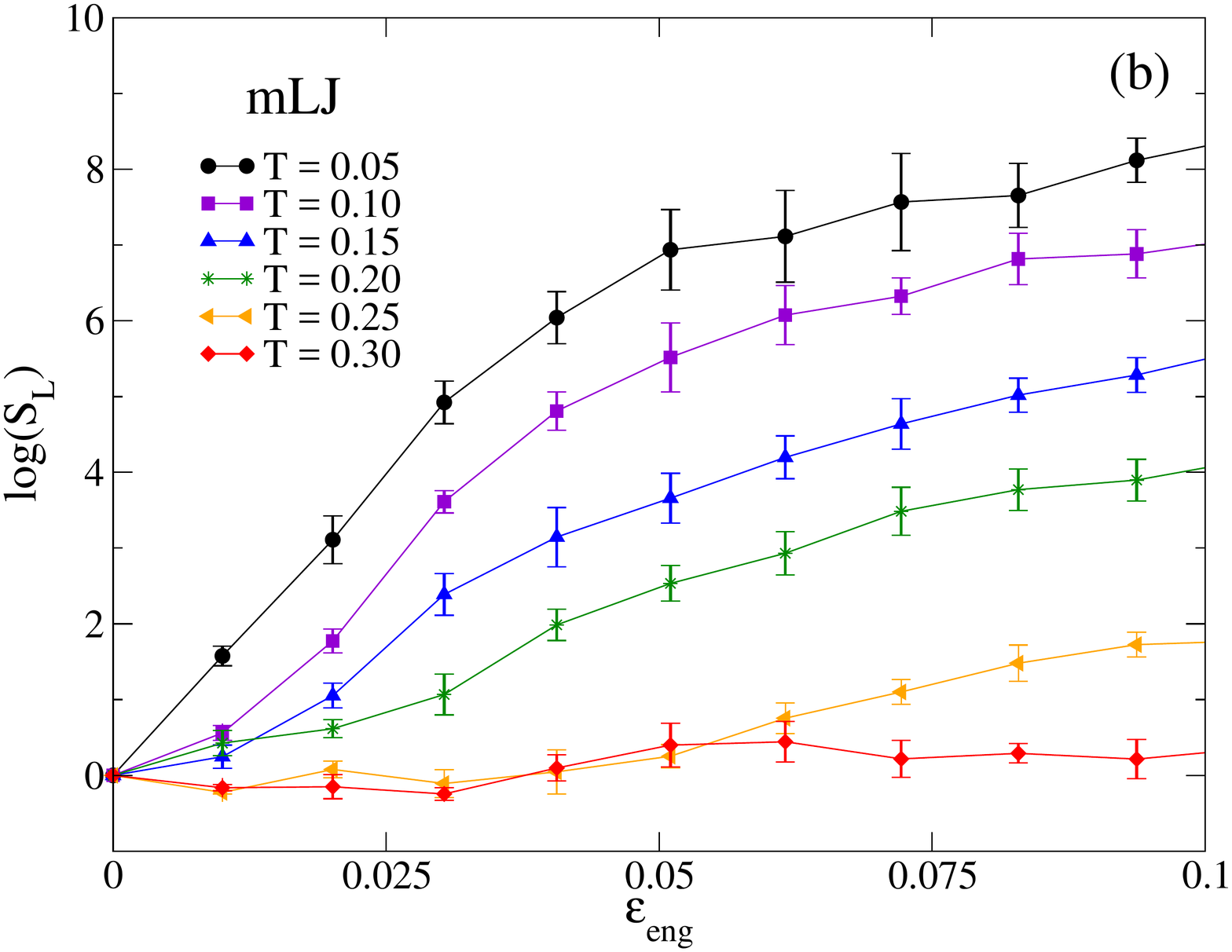}
			\end{subfigure}
			\caption{\footnotesize{Spatial fluctuation of local strain rate ($S_L$) as a function of strain for (a) LJ and (b) mLJ pillars deformed at temperatures below $T_g$ using uniaxial deformation rate of $\dot{\epsilon} = 1\times10^{-4}$. $S_L$ is calculated using a time window that corresponds to $\epsilon = 1\%$, and each pillar is divided into 20 slabs in length ($z$) such that each slab has a thickness of $5.0 ~\sigma$ to $5.5 ~\sigma$, before and after 30\% strain is applied, respectively. }}
			\label{fig:Sl}
		\end{figure}

\begin{figure}[H]
			\centering
			\includegraphics[scale = 0.4]{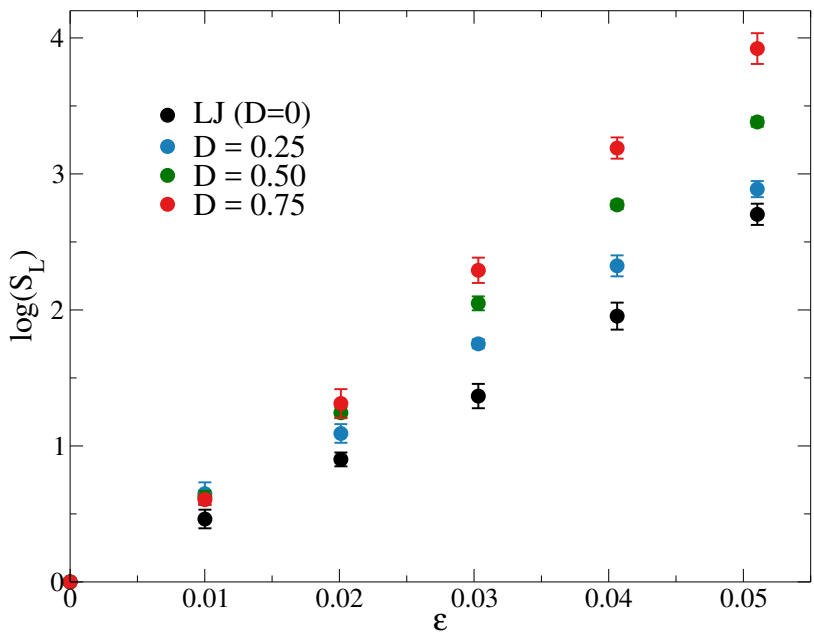}
			\caption{\footnotesize{$S_L$ as a function of strain for each D value at $T = 0.05$. $J_2$ values used to compute $S_L$ here are cumulative, i.e., between the current strained configuration and initial configuration. }}
			\label{fig:sl-D}
		\end{figure}

We quantify the extent of strain localization by examining the $S_L$ function defined above for all of the systems as a function of strain. In Figure \ref{fig:Sl}, we plotted $\log S_L$ as a function of total strain. In both the LJ and the mLJ systems, $\log S_L$ at a given strain increases sharply as temperature decreases, and this increase is larger in the mLJ system. At higher temperatures, the $\log S_L$ values remain relatively unchanged throughout deformation, consistent with the snapshots shown in Figure \ref{fig:pillars} and the expectation for a homogeneous strain field. At lower temperatures, the magnitude of the spatial variations in strain rate becomes substantially larger as strain increases due to the formation of either a neck or the shear band. As $T$ is lowered, $S_L$ gradually increases with no sharp cross-over from homogeneous to localized plastic flow. Figure \ref{fig:sl-D} shows that for different values of $D$, the $S_L$ values differ by up to one order of magnitude between the two extreme cases of $D$ values as the sample is deformed through the yield point. While $S_L$ does distinguish homogeneous dynamics from strain localization, $S_L$ does not allow us to draw a distinctive boundary between necking and brittle failures. 
	
\begin{figure}[H]
			\centering
			\begin{subfigure}{0.45 \columnwidth}
				\includegraphics[width = \linewidth]{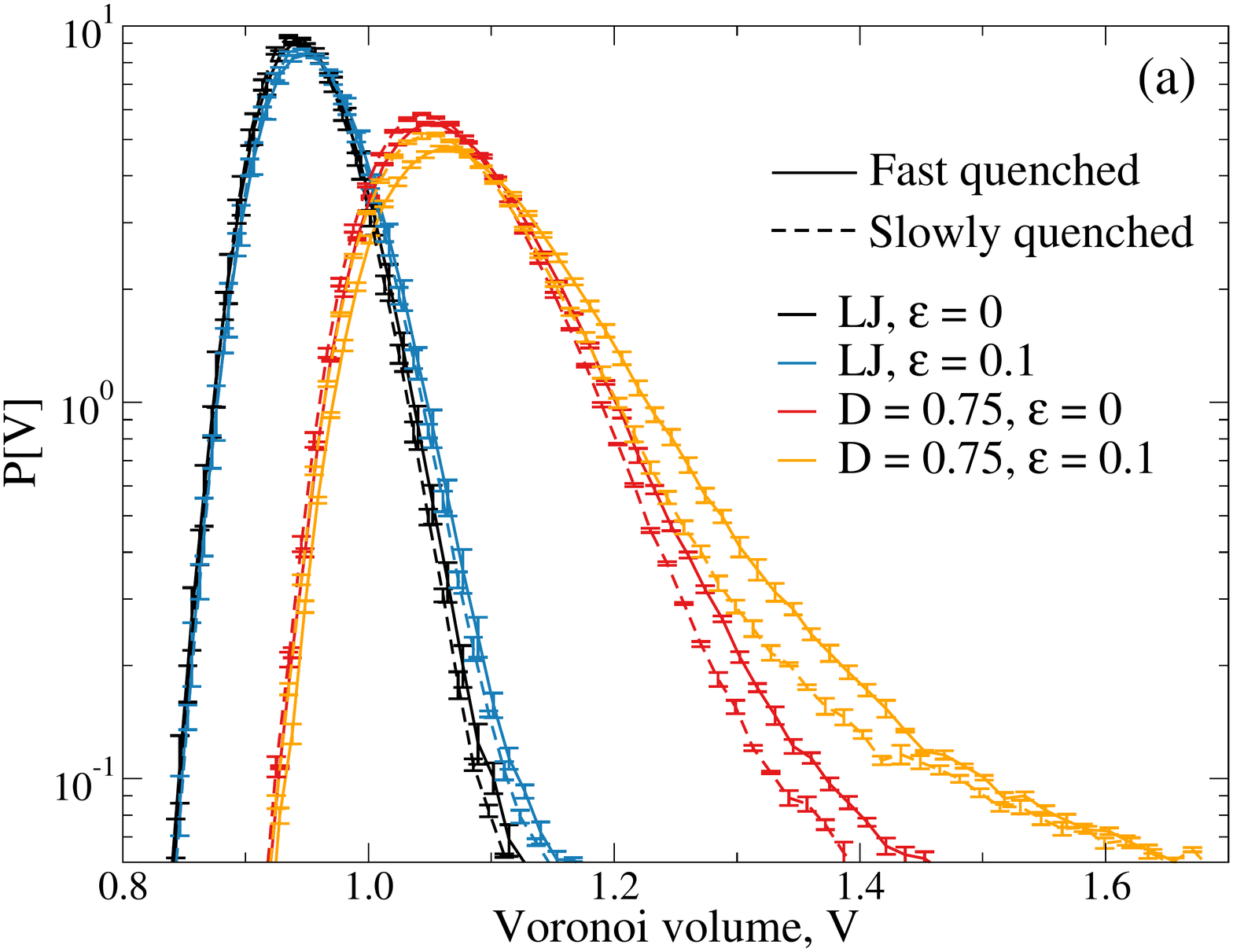}
			\end{subfigure}
			~
			\begin{subfigure}{0.45 \columnwidth}
				\includegraphics[width = \linewidth]{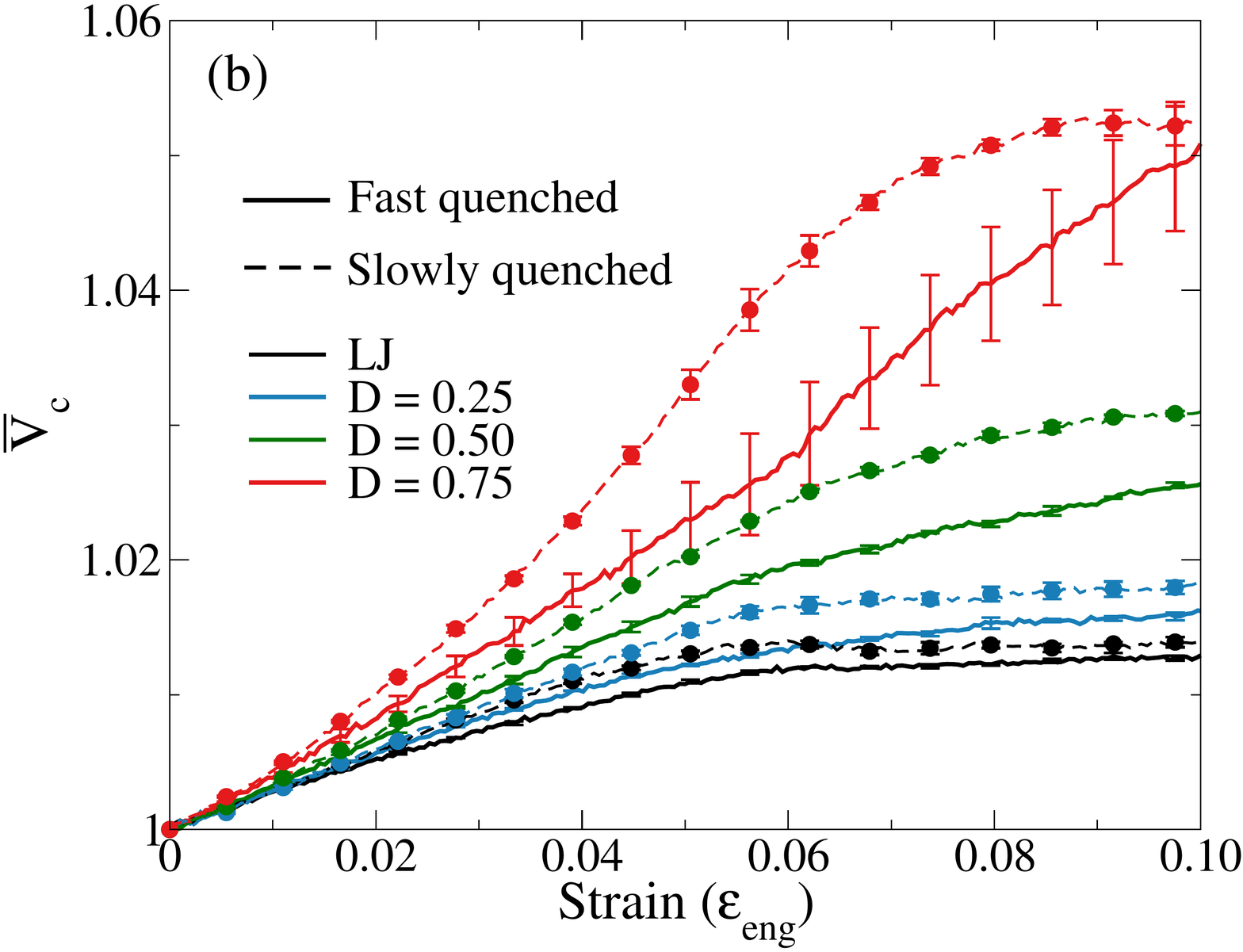}
			\end{subfigure}
			\caption{\footnotesize{(a). Voronoi volume distributions for LJ and $D = 0.75$ before deformation and after yield. (b). Cutoff Voronoi volume for the four D values tested in our study, at T = 0.05, on samples generated by using different quench rates. Solid lines represent fast quenched samples ($\dot{\Gamma} = 1\times10^{-3}$), while dashed lines represent slowly quenched samples ($\dot{\Gamma} = 1\times10^{-5}$). Error bars for all samples are generated using standard errors over three independent configurations.}}
			\label{fig:voro}
		\end{figure}

To distinguish between ductile necking and brittle fracture, we examine the Voronoi volumes for the monomers in our nanopillars. When we plot the Voronoi volume distributions for the samples experiencing brittle and ductile failure modes, we observe that the brittle samples have significant changes in Voronoi volume distributions after strain is applied. Figure \ref{fig:voro}a shows that for the most brittle nanopillar (mLJ, $D = 0.75$), the Voronoi volume distribution after yield is characterized by a broad tail that favors larger Voronoi cell volumes. This change is not observed in the LJ potential systems ($D = 0$) after the same amount of strain is applied. Additionally, at the highest temperature ($T = 0.3$), the difference in the distributions before and after strain is insignificant (data not shown), as expected for a ductile system. From this observation, we devised a simple metric to differentiate between the failure modes: for each applied strain, we identify the 90th percentile Voronoi volume, and refer to this value as the cutoff volume,  $V_c(\epsilon)$. We then plotted this cutoff volume as a function of strain for the various D parameters we have used in our study; to better compare these cutoff volumes across different systems, we normalized $V_c$ at each strain by its value before deformation, $\bar{V}_c = V_c (\epsilon) / V_c (\epsilon = 0)$. 

Figure \ref{fig:voro}b shows the cutoff volume vs. strain for all four D parameters used, at the lowest temperature we have studied (T = 0.05). Because of the normalization procedure, the $\bar{V}_c$ values for all systems start at 1, but the mLJ systems with higher D values increase more rapidly than the LJ systems after a small strain is applied, prior to when the yield point is observed in macroscopic analyses of the system (i.e. stress-strain curves, pillar snapshots). For the pillars with more ductile response, $\bar{V}$ exhibits a modest increase of approximately 1\% while the more brittle pillars exhibit a more rapid and larger increase, up to approximately 5\%. For more slowly quenched pillars, $\bar{V}_c$ values are higher at a given strain for each $D$, suggesting that the slowly quenched samples are embrittled due to their more rapid increase in the particles' relative free volume.

\subsection{Distinguishing Failure Modes}

The mechanical properties of glass are highly dependent on surface treatment, and previous studies have argued that the response of different materials can be understood through changes in surface roughness.\cite{Nie2009-surface-treatment, Srolovitz-Greer2015} Here we attempt to quantify whether our observed failure modes correlate with changes in surface roughness. From our pre-deformation configurations, we calculated the surface roughness as $\sigma^2_R = \sum_{i=1}^{n} \left(R_i - \left<  R \right>_{pillar}\right)^2$, where we summed the squared differences between $i^{th}$ slab radius, $R_i$, and the averaged pillar radius, $\left<R\right>_{pillar}$, for all of our pillars, and plotted them against the fluctuation in spatial correction of local strain rate, $S_L$. As demonstrated above in Figure \ref{fig:Sl}, lower temperature and higher $D$ values result in higher $S_L$ values. For systems with $D > 0$, the pillars tend to have a rougher surface. However, the variation in roughness is not significant between samples with the same $D$ value and temperature, even when we compare the coldest pillar with the hottest pillar, which clearly have different failure modes and $S_L$ values. This result suggests that surface roughness alone cannot adequately differentiate the different degrees of strain localization.  A similar lack of correlation is also observed between $\bar V_c$ and the surface roughness.

		\begin{figure}[H]
			\centering
				\includegraphics[scale = 0.4]{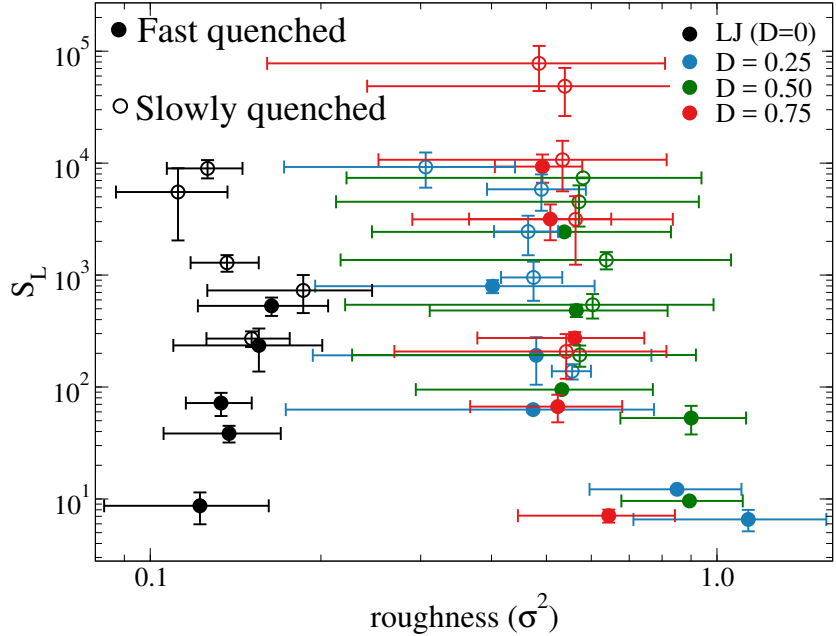}
			\caption{\footnotesize{Spatial fluctuation of local strain rate as a function of pillar surface roughness for fast (closed circles) and slowly (open circles) quenched pillars in our studies. Each data point is an average of three independent configurations.}}
			\label{fig:roughness}
		\end{figure}

\section{Summary}

We used Molecular Dynamics (MD) systems of glassy polymer nanopillars to simulate uniaxial tensile deformation. In order to simulate brittle failure, we incorporated a previously proposed change to the Lennard-Jones (LJ) pair potential and further characterized the impact of this modification on macroscopic material properties as well as structural and dynamic information on the length scale of merely a few Kuhn lengths. Using an order parameter that quantifies the localized strain rate spatial fluctuation, we found that the mLJ potential can be used to describe significant differences in the degree of strain localization as a function of D values and temperature, which allows it to differentiate from homogeneous plastic flow in the cases when no shear band forms. From calculating the Voronoi volumes of each monomer, we found that the brittle nanopillars have drastically different Voronoi volume distributions from the ductile samples, and we devised a cutoff Voronoi volume at 90th percentile to quantitatively characterize the nanopillars with different failure modes. We believe this metric would enable us to quantitatively understand different failure modes using features in the sample structure at the molecular scale. 

\section{Acknowledgement}

The authors would like to gratefully acknowledge our funding sources: National Science Foundation (NSF) Civil, Mechanical and Manufacturing Innovation (CMMI15-3691), partial support from Materials Research and Engineering Center (MRSEC) at the University of Pennsylvania (DMR-1720530). This work made use of computational resources provided through NSF Extreme Science and Engineering Discovery Environment (XSEDE) award DMR-150034. We also thank Robert J.~S.~Ivancic for helpful discussions. 

\bibliography{Library}
\end{document}